---------------------------------------------------------------------------
############################## RevTex Copy #################################
---------------------------------------------------------------------------

\documentstyle[aps]{revtex}

\newcommand{\bel}[1]{\begin{equation}\label{#1}}
\newcommand{\be}{\begin{equation}}
\newcommand{\ee}{\end{equation}}

\newcommand{\beal}[1]{\begin{eqnarray}\label{#1}}
\newcommand{\bea}{\begin{eqnarray}}
\newcommand{\eea}{\end{eqnarray}}

\newcommand{\bean}{\begin{eqnarray*}}
\newcommand{\eean}{\end{eqnarray*}}

\newcommand{\ba}{\begin{array}}
\newcommand{\ea}{\end{array}}

\newcommand{\bab}{\begin{abstract}}
\newcommand{\eab}{\end{abstract}}

\newcommand{\bml}{\begin{mathletters}}
\newcommand{\eml}{\end{mathletters}}

\newcommand{\q}{\quad}
\newcommand{\qq}{\quad\quad}

\newcommand{\bfm}[1]{\mbox{\boldmath $#1$}}

\newcommand{\dv}{\partial}

\newcommand{\bad}[2]{\left( \begin{array}{c}{#1}\\{#2}\end{array} \right)} 
\newcommand{\bat}[3]{\left( \begin{array}{c}{#1}\\{#2}\\
{#3}\end{array} \right)}

\newcommand{\bamd}[4]{\left( \begin{array}{cc}{#1}&{#2}\\
{#3}&{#4}\end{array} \right)}
\newcommand{\bamt}[9]{\left( \begin{array}{ccc}{#1}&{#2}&{#3}\\{#4}&{#5}&{#6}\\
{#7}&{#8}&{#9}\end{array} \right)}

\newcommand{\bamq}[4]{\left( \begin{array}{cccc}{#1}&{#2}&{#3}&{#4}\\}
\newcommand{\bamc}[5]{\left( \begin{array}{ccccc}{#1}&{#2}&{#3}&{#4}&{#5}\\}
\newcommand{\eam}{\end{array} \right)}

\newcommand{\law}{\leftarrow}

\newcommand{\raw}{\rightarrow}

\newcommand{\lLRw}{\Longleftrightarrow}

\newcommand{\ag}{\alpha}
\newcommand{\bg}{\beta}
\newcommand{\cg}{\gamma}

\newcommand{\lgg}{\lambda}

\newcommand{\sg}{\sigma}

\newcommand{\fs}{\footnotesize}

\newcommand{\bii}{\begin{itemize}}
\newcommand{\eii}{\end{itemize}}

\newcommand{\ben}{\begin{enumerate}}
\newcommand{\een}{\end{enumerate}}

\newcommand{\bq}{\begin{quote}}
\newcommand{\eq}{\end{quote}}

\newcommand{\bc}{\begin{center}}
\newcommand{\ec}{\end{center}}

\newcommand{\btb}{\begin{table}}
\newcommand{\etb}{\end{table}}

\newcommand{\bt}{\begin{tabular}}
\newcommand{\et}{\end{tabular}}

\newcommand{\br}{\begin{flushright}}
\newcommand{\er}{\end{flushright}}

\newcommand{\bl}{\begin{flushleft}}
\newcommand{\el}{\end{flushleft}}

\newcommand{\th}[1]{\thanks{#1}}

\newcommand{\f}[1]{\footnote{#1}}

\newcommand{\bref}{\begin{references}}
\newcommand{\eref}{\end{references}}

\newcommand{\bb}{\begin{thebibliography}{99}}
\newcommand{\eb}{\end{thebibliography}}
\newcommand{\bi}{\bibitem}

\newcommand{\btp}{\begin{titlepage}}
\newcommand{\etp}{\end{titlepage}}

\newcommand{\go}{\section{Introduction}}

\newcommand{\axp}[3]{Ann.~Phys.~(NY)                    {\bf #1},  #2  (19#3)}

\newcommand{\ixa}[3]{Int.~J.~Mod.~Phys.~A               {\bf #1},  #2  (19#3)}

\newcommand{\jxg}[3]{J.~Phys.~A                         {\bf #1},  #2  (19#3)}

\newcommand{\mxb}[3]{Mod.~Phys.~Lett.~A                 {\bf #1},  #2  (19#3)}

\newcommand{\nxb}[3]{Nucl.~Phys.                        {\bf #1},  #2  (19#3)}

\newcommand{\pxi}[3]{Phys.~Lett.                        {\bf #1},  #2  (19#3)}

\newcommand{\pxxa}[3]{Prog.~Theor.~Phys.                {\bf #1},  #2  (19#3)}

\newcommand{\xxx}[3]{{\bf #1},  #2  (19#3)}

\title{HALF-WHOLE DIMENSIONS\\ IN\\ QUATERNIONIC QUANTUM MECHANICS}

\author{Stefano De Leo\th{{\sl deleos@le.infn.it}}}
\address{Dipartimento di Fisica - Istituto Nazionale di Fisica Nucleare\\
- Lecce, 73100, Italy -}

\date{\today}

\draft

\begin{document}

\maketitle

\bab
We introduce {\em half-whole} dimensions for quaternionic matrices and 
propose a quaternionic version of the Frobenius-Schur theorem which allows 
us to obtain the proper quaternionic dimensionality for the representations 
of the Dirac and Duffin-Kemmer-Petiau (DKP) algebras.
\eab


\renewcommand{\thefootnote}{\sharp\arabic{footnote}}

\go

We briefly recall the main properties of the quaternionic field. Such a 
field is characterized by three imaginary units $i$, $j$, 
$k$ which satisfy the following multiplication rules:
\bml
\be
i^{2}=j^{2}=k^{2}=-1 \q ,
\ee
\be
[i, \; j]=2k \q , \q [j, \; k]=2i \q , \q [k, \; i]=2j \q ,
\ee
\eml
In going from the complex numbers to the quaternions we lose the 
property of the commutativity. The {\em full}-quaternionic conjugation is 
denoted by $\dag$ and defined by
\[ 1^{\dag} = 1 \q , 
\q (i, \; j, \; k)^{\dag}= - ~ (i, \; j, \; k) \q. \]
The previous definition implies
\[ (\psi \phi)^{\dag}=\phi^{\dag} \psi^{\dag} \q ,\]
for $\psi$, $\phi$ quaternionic functions.

Working in quaternionic quantum mechanics with quaternionic geometry 
($QQM_{qg}$) there is no quaternionic self-adjoint operator with all the 
properties expected for a momentum operator (\cite{adl}, pag.~63). 
We like overcoming such a difficulty 
using a complex scalar product~\cite{hor} (or complex geometry as called by 
Rembieli\'nski~\cite{rem})
\[ <\psi \mid \phi>_{c} \; = \frac{1}{2} \; 
( \; <\psi \mid \phi> - \; i <\psi \mid \phi> i \; ) \q , \]
and defining as the appropriate momentum operator~\cite{rot}
\bel{bar}
{\bf p}\equiv -\bfm{\dv} \mid i \qq ({\bf p}\psi\equiv -\bfm{\dv} \psi i) 
\q .
\ee
Note that the usual ${\bf p}\equiv -i\bfm{\dv}$, still gives a self-adjoint 
operator with standard commutation relations with the coordinates, but such an 
operator does not commute with the Hamiltonian, which will be, in general, a 
quaternionic quantity. 

In eq.~(\ref{bar}), a particular {\em barred} operator appears.  
We recall the {barred} quaternion definition (for further details, see 
ref.~\cite{del1}):
\[ (q+p\mid i)r \equiv qr + pri \qq [~q, \; p, \; r \in {\cal H}~] \q . \]

We observe that the dimensionality of a complete set of states for complex 
inner product $<\psi \mid \phi>_{c}$ is {\em twice} 
that for the quaternionic inner product $<\psi \mid \phi>$. 
Specifically, if $\mid \eta_{m}>$ represent a complete set of intermediate 
states for 
the quaternionic scalar product, so that
\[ <\psi \mid \phi> \; = \sum_{m} <\psi \mid \eta_{m}><\eta_{m} \mid \phi> \; 
\; ,\]
$\mid \eta_{m}>$ and  $\mid \eta_{m} j>$ form a complete set of states for 
the complex scalar product,
\bean 
\mid \phi> & = & \sum_{m} \;  ( \; 
\mid \eta_{m}><\eta_{m} \mid \phi>_{c}+ 
\mid \eta_{m}j><\eta_{m} j\mid \phi>_{c}\\
 & = & \sum_{n} \mid \chi_{n}><\chi_{n} \mid \phi>_{c} \; \; ,
\eean
where $\chi_{n}$ represent {\em complex} orthogonal states. The  
completeness relations can be written as\f{For further details on these  
completeness relations, the redear can consult the interesting work of 
Horwitz and Biedenharn, cited in ref.~\cite{hor}, pag.~455.}
\bean  
\stackrel{\raw}{\bf 1} & = & \sum_{n} \mid \chi_{n}>( \; \chi_{n} \mid \; \; ,\\
\stackrel{\law}{\bf 1} & = & \sum_{n} \mid \chi_{n} \; )<\chi_{n} \mid \; \; ,
\eean
where, the standard Dirac's notation is generalized by the following 
definitions
\bean ( \; \chi_{n} \mid \phi> & = & 
<\chi_{n} \mid \phi>_{c} \q ,\\
<\phi \mid \chi_{n} \; ) & = & 
<\phi \mid \chi_{n} >_{c} \q .
\eean

\section{Even dimensions}

Within quaternionic quantum mechanics with complex geometry ($QQM_{cg}$) we 
can introduce  a ``new'' complex-imaginary unit 
\[ 1\mid i \qq \qq [~(1\mid i)\psi \equiv \psi i~] \q , \] 
\[ (\; 1\mid i \; )^{2}=-1 \q , \q (\; 1\mid i \; )^{\dag}=-1\mid i \q ,\]
which commutes with $i, \; j, \; k$. In order to prove the antihermiticity 
of $1\mid i$,  we note that with complex scalar products we have
\[ <\psi \mid \phi i>_{c} ~~ = ~~ <\psi \mid \phi>_{c} \; i ~~=~~ i 
<\psi \mid \phi>_{c} ~~=~~  - <\psi i \mid \phi >_{c} \q . \]
Thanks to this ``new'' complex-imaginary 
unit we can perform a translation between even-dimensional 
complex matrices and quaternionic matrices with half the 
dimensions~\cite{del1}. Working in $QQM_{cg}$, a generic $2n\times 2n$ 
complex representation $M$ 
can be reduced to two $n$-dimensional quaternionic representations 
$M_{1}$ and $M_{2}$
\be
M=M_{1} \oplus M_{2} \q .
\ee
We give the explicit construction that establishes reducibility for the 
case of $2 \times 2$ complex matrices
\be
M = \bamd{c_{1}}{c_{2}}{c_{3}}{c_{4}} \q .
\ee
As consequence of our complex geometry we have a {\em doubling} of states:
\be
\psi = \bad{1}{0} \; , \; \bad{0}{1} \q \mbox{and} \q 
\bad{j}{0} \; , \; \bad{0}{j} \q .
\ee
We observe that the last two ($j$-complex) states cannot mix, under the 
action of $M$, with the 
former because of the complex nature of the 2-dimensional complex matrix 
$M$. Thus, the vector space is {\em reducible}. Requiring the following 
transformation for the previous states
\be
\tilde{\psi} = S \psi \q ,
\ee
with respectively
\be
\tilde{\psi} = \bad{1}{0} \; , \; \bad{j}{0} \q \mbox{and} \q 
\bad{0}{1} \; , 
\; \bad{0}{j} \q ,
\ee
we can quickly find the quaternionic similarity matrix $S$ 
($S^{\dag}$=$S^{-1}$) which reduces the complex representation, $M$. 
Explicitly, we have
\be
S = \bamd{a}{~~ja}{-jd}{~d} \qq [~S^{\dag}=S~]  \q ,
\ee
with
\bc
\bt{lcc}
$2a=1-i\mid i$ &~~~which extinguishes $j$-complex elements~~~ & ,\\ 
$2d=1+i\mid i$ &~~~which extinguishes complex elements~~~ & . 
\et
\ec
The transformed matrix $\tilde{M}=SMS^{\dag}$ is then given by
\be
\tilde{M} = \bamd{q_{1}+p_{1}\mid i}{0}{~0}{~~q_{2}+p_{2}\mid i} \q ,
\ee
where
\bean
2q_{1} & = & c_{1}+c_{4}^{*}+j(c_{3} -c_{2}^{*}) \q ,\\
2ip_{1} & = & c_{1}-c_{4}^{*}-j(c_{3}+c_{2}^{*})  \q ,\\
2q_{2} & = & c_{1}^{*}+c_{4}+j(c_{3}^{*} -c_{2})\q ,\\
2ip_{2} & = & c_{1}^{*}-c_{4}-j(c_{3}^{*}+c_{2}) \q .
\eean
Thanks to this reduction we can obtain a set of rules for the translation. 
The already well known identifications of $i$, $j$ and $k$ with 
$-\frac{i}{2}\bfm{\sg}$ (\bfm{\sg} the Pauli matrices), and of course 
$1$ (in $\cal H$) with the $2$-dimensional unit matrix, can thus be extended to 
the most general $2$-dimensional complex matrix.
\bml
\bea
M=\bamd{c_{1}}{c_{2}}{c_{3}}{c_{4}} & \q \lLRw \q & 
M_{1}=q_{1} + p_{1} \mid i\\ 
M^{*}=\bamd{c_{1}^{*}}{c_{2}^{*}}{c_{3}^{*}}{c_{4}^{*}} & \q \lLRw \q & 
M_{2}=q_{2} + p_{2} \mid i
\eea
\eml
\[ [~c_{1, \; ..., \; 4} \in {\cal C}(1, \; i) \; \; \mbox{and} \; \;
q_{1, \; 2} \; , \; p_{1, \; 2}  \in {\cal H}~] \q . \]

Obviously we can generalize the previous result for a generic 
$2n$-dimensional complex matrix. In particular, $4\times 4$ complex matrices 
(with four {\em complex} states) split into $2\times 2$ quaternionic 
matrices (with two {\em complex} $+$ two {\em $j$-complex} states): 
\beal{4}
M=\bamq{c_{1}}{c_{2}}{c_{3}}{c_{4}} c_{5} & c_{6} & c_{7} & c_{8} \\
c_{9} & c_{10} & c_{11} & c_{12} \\
c_{13} & c_{14} & c_{15} & c_{16} 
\eam & \q \lLRw \q & 
M=\bamd{r_{1} + s_{1} \mid i}{~~r_{2} + s_{2} \mid i}
{r_{3} + s_{3} \mid i}{~~r_{4} + s_{4} \mid i}
\eea
\[ [~c_{1, \; ..., \; 16} \in {\cal C}(1, \; i) \; \; \mbox{and} \; \;
r_{1, \; ..., \; 4} \; , \; s_{1, \; ..., \; 4}  \in {\cal H}~] \q ,\]
where
\bean
2r_{1} & = & c_{1}+c_{6}^{*}+j(c_{5} -c_{2}^{*}) \q ,\\
2is_{1} & = & c_{1}-c_{6}^{*}-j(c_{5}+c_{2}^{*})  \q ,\\
2r_{2} & = & c_{3}+c_{8}^{*}+j(c_{3} -c_{8}^{*}) \q ,\\
2is_{2} & = & c_{7}-c_{4}^{*}-j(c_{7}+c_{4}^{*})  \q ,\\
2r_{3} & = & c_{9}+c_{14}^{*}+j(c_{13} -c_{10}^{*}) \q ,\\
2is_{3} & = & c_{9}-c_{14}^{*}-j(c_{13}+c_{10}^{*})  \q ,\\
2r_{4} & = & c_{11}+c_{16}^{*}+j(c_{15} -c_{12}^{*}) \q ,\\
2is_{4} & = & c_{11}-c_{16}^{*}-j(c_{15}+c_{12}^{*})  \q.
\eean

\section{Odd Dimensions}

As described in a recent article~\cite{del2} the above translation can be 
performed, 
using a particular trick, for odd dimensional complex representations. 
$3\times 3$ complex matrices can be reduced to two overlapping 
$2\times 2$ block forms (so that the $(2,2)$-element is common to both 
blocks). We start with a generic $3\times 3$ complex matrix
\be
M = 
\bamt{c_{1}}{c_{2}}{c_{3}}{c_{4}}{c_{5}}{c_{6}}{c_{7}}{c_{8}}{c_{9}} \q ,
\ee
which shows the following {\em doubling} of base states, in the associated 
vector space:
\[ \bat{1}{0}{0} \; , \; \bat{0}{1}{0} \; , \; \bat{0}{0}{1} \q \mbox{and}
\q \bat{j}{0}{0} \; , \; \bat{0}{j}{0} \; , \; \bat{0}{0}{j} \q .\] 
As remarked in section II, the vector space is {\em reducible}. The 
quaternionic similarity matrix $S$ which transforms the previous states in
\[ \bat{1}{0}{0} \; , \; \bat{j}{0}{0} \; , \; \bat{0}{1}{0} \q \mbox{and}
\q \bat{0}{0}{1} \; , \; \bat{0}{0}{j} \; , \; \bat{0}{j}{0} \q ,\] 
and performs the reduction is
\bml
\bea
S & = & \bamt{a}{~ja~}{0~}{0}{0}{1~}{-jd}{d}{0~} \q ,\\
\nonumber  & & \\
S^{\dag} & = & \bamt{a}{0}{~ja~}{~-jd~}{0}{d}{0}{1}{0} \q .
\eea
\eml
The transformed matrix $\tilde{M}$ is then given by
\be
\tilde{M} = \bamt{~(c_{1}+jc_{4})a+(c_{5}^{*}-jc_{2}^{*})d}
{~~~~(c_{3}+jc_{6})a~~~~}
{~~0}{~c_{7}a-jc_{8}^{*}d}{~~~~c_{9}~~~~}{~~jc_{7}^{*}a+c_{8}d}{0}
{~~~~(-jc_{3}+c_{6})d~~~~}
{~~~~(c_{1}^{*}+jc_{4}^{*})a+(c_{5}-jc_{2})d~} \q .
\ee
In $\tilde{M}$ the $(2,2)$-element can be written conveniently 
as $c_{9}(a+d)$, i.~e. containing a sum of projection operators.

Thus, we can translate a generic $3\times 3$ complex matrix by a 
{\em particular} $2\times 2$ quaternionic matrix
\bml
\beal{3}
M=
\bamt{c_{1}}{c_{2}}{c_{3}}{c_{4}}{c_{5}}{c_{6}}{c_{7}}{c_{8}}{c_{9}} &
\q \lLRw \q & M_{1} = \bamd{~(c_{1}+jc_{4})a+(c_{5}^{*}-jc_{2}^{*})d}
{~~~~(c_{3}+jc_{6})a~}{~c_{7}a-jc_{8}^{*}d}{~~~~c_{9}a~}\\
M^{*}=
\bamt{c_{1}^{*}}{c_{2}^{*}}{c_{3}^{*}}{c_{4}^{*}}{c_{5}^{*}}{c_{6}^{*}}
{c_{7}^{*}}{c_{8}^{*}}{c_{9}^{*}} & \q \lLRw \q &
M_{2}=\bamd{~c_{9}d}{~~~~jc_{7}^{*}a+c_{8}d~}{~(-jc_{3}+c_{6})d}
{~~~~(c_{1}^{*}+jc_{4}^{*})a+(c_{5}-jc_{2})d~} \q .
\eea
\eml
In order to prove the last identification, note that
\be
\bamd{0}{1}{-j}{0} \; M_{2} \; 
\bamd{0}{j}{1}{0} =  
\bamd{~(c_{1}^{*}+jc_{4}^{*})a+(c_{5}-jc_{2})d}
{~~~~(c_{3}^{*}+jc_{6}^{*})a~}{~c_{7}^{*}a-jc_{8}d}{~~~~c_{9}^{*}a~} \q .
\ee
We conclude this section, remarking the difference between a 
$2\times 2$ quaternionic matrix 
which acts non-trivially only on three states [see eq.~(\ref{3})]
\be
M^{(\mbox{{\fs three}})}=\bamd{~q+p\mid i}{~~~~ra~}{~z_{1}a+jz_{2}d}
{~~~~z_{3}a~}
\qq [~z_{1, \; 2, \; 3} \in {\cal C}(1, \; i) \; \; \mbox{and} \; \;
q, \; p, \; r  \in {\cal H}~] \q ,
\ee
\bc
-- $M^{(\mbox{{\fs three}})}$ action --
\ec
\[ M^{(\mbox{{\fs three}})}\bad{0}{jz}=\bad{0}{0} \; , \;
M^{(\mbox{{\fs three}})}\bad{q}{z}=\bad{q'}{z'} \qq 
[~z, z' \in {\cal C}(1, \; i) \; \; \mbox{and} \; \;
q, \; q' \in {\cal H}~] \q ,\]
and a generic $2\times 2$ quaternionic matrix which acts non trivially on 
four states [see eq.~(\ref{4})].

In standard theory the dimensionality of complex matrices is strictly 
connected to the dimensionality of the vector space, whereas working in 
$QQM_{cg}$ we have a {\em doubling} of states, so we require the following 
correspondence rule between the dimensionality of quaternionic matrices 
($n$) and the dimensionality of the vector space ($n_{\mbox{vs}}$)
\[ 2n=n_{\mbox{vs}} \q .\]
In order to distinguish between {\em odd} and {\em even} vector spaces we 
introduce {\em half-whole} dimensions for our quaternionic matrices
\be
n_{M^{(\mbox{{\fs three}})}}=\frac{3}{2} \q , \q 
n_{M^{(\mbox{{\fs four}})}}=2  \q .
\ee

\section{Dirac Algebra}
Let us consider {\em in abstracto}, four algebraic quantities $\cg^{\mu}$, 
which satisfy the Dirac relations
\be \cg^{\mu} \cg^{\nu} + \cg^{\nu} \cg^{\mu} = 2 g^{\mu\nu} \qq
(\mu , \; \nu = 0, \; 1, \; 2, \; 3) \q . \ee  
We observe that in the following array (characterized by sixteen 
quantities)
\bc
\bt{l}
$1 \q ;$\\ \\
$\cg^{0} \; , \; \cg^{1} \; , \; \cg^{2} \; , \; \cg^{3} \q ;$\\ \\
$\cg^{0}\cg^{1} \; , \; \cg^{0}\cg^{2} \; , \; \cg^{0}\cg^{3} \; , \; 
\cg^{1}\cg^{2} \; , \;\cg^{1}\cg^{3} \; , \; \cg^{2}\cg^{3} \q ;$\\ \\
$\cg^{0}\cg^{1}\cg^{2} \; , \; \cg^{0}\cg^{1}\cg^{3} \; , \; 
\cg^{0}\cg^{2}\cg^{3} \; , \; \cg^{1}\cg^{2}\cg^{3} \q ;$\\ \\
$\cg^{0}\cg^{1}\cg^{2}\cg^{3} \q ;$
\et
\ec
any product of two elements is proportional to another element of the array. 
We now wish to obtain appropriate matrix representations for the abstract 
algebraic quantities $\cg^{\mu}$. First of all we briefly recall the 
standard (complex) results. After that we will generalize our 
considerations, by considering, as underlying numerical fields, quaternions 
and complexified quaternions.

In standard (complex) theory it is very simple to prove the following 
theorems\f{In order to simplify next considerations we indicate by 
$\cg_{A}$ $(A=1, \; 2, \; ... \; , 16)$ the general element of the array.}.

1 -- If $\cg_{A}\neq 1$, one can always find a $\cg_{B}$ such that 
$\cg_{B}\cg_{A}\cg_{B}=-\cg_{A}$;

2 -- With the exception of the 1--element, the trace of all $\cg_{A}$'s is 
zero;

3 -- The sixteen $\cg_{A}$'s are linearly independent,
\[ \sum_{A=1}^{16} \ag_{A} \cg_{A} = 0 \qq (\ag_{A} \; \; 
\mbox{complex numbers}) \]
if and only if all the sixteen coefficients $\ag_{A}$ vanish;

4 -- The only hypercomplex quantity $X=\sum_{A=1}^{16} \ag_{A} \cg_{A}$ 
which commutes with all $\cg_{A}$'s is (a multiple of) the unity.

In order to find all possible irreducible representations of the Dirac 
algebra, we shall need two remarkable theorems regarding the 
representations of algebras. 

The first is the theorem of Frobenius and Shur 
which may stated as follows: 

5 -- Let $\cal A$ be an algebra of order $n$ 
possessing a unit element. Let $p$ be the number of (non-equivalent) 
irreducible representations of the algebra, and denote the dimensionality 
of these representations by $n_{1}, \; n_{2}, \; ... \; ,n_{p}$ in turn. 
Then
\be n=n_{1}^{ \; 2}+n_{2}^{ \; 2}+ \; ... \; +n_{p}^{ \; 2} \q . \ee

The second theorem enables to find the number $p$ of the possible 
irreducible representations: 

6 -- If the algebra $\cal A$ is semi-simple, then 
the number of possible irreducible representations is equal to the maximum 
number of base elements which commute with each other. 

Combining this two theorems (5-6), we can quickly obtain the dimensionalities 
of the various possible irreducible representations of a semi-simple algebra 
with a unit element.

In virtue of previous considerations one finds that the only {\em complex} 
irreducible representations of the Dirac algebra is {\em four-dimensional}.

What happens for quaternions? Obviously the 
theorems 1 and 6 also hold, since their demonstration don't use the 
explicit form of the $\cg_{A}$-matrices. In order to prove theorems 2, 3, 4 
we must introduce an appropriate definition of trace and choose 
{\em commuting} numerical coefficients $\ag_{A}$. Finally the remaining 
(Frobenius and Shur) theorem will be timely modified.

\section{Quaternionic Dirac Algebra}
In a previous work, Rotelli~\cite{rot} derived a {\em new} version of the Dirac 
equation by adopting quaternions as underlying numerical field. The main 
difference between quaternionic and complex Dirac equation is represented 
by the dimensionality of the $\cg^{\mu}$-matrices. In fact, working within 
QQM, there exists a $2\times 2$ matrix representation for the Dirac algebra 
given by
\bel{rm}
\bfm{\cg} = {\bf Q} \bamd{0}{1}{1}{0} \q , \q \cg^{0} = \bamd{1}{0}{0}{-1}
\qq [{\bf Q}\equiv (i, \; j, \; k)] \q .
\ee
Notwithstanding the two component structure of the quaternionic wave 
functions, four standard Dirac solutions are reproduced. In such an 
equation, the 
complex geometry gives a welcome doubling of 
states\f{Observe that within QQM with complex geometry 
$e^{-ipx}, \; je^{-ipx}$ represent orthogonal solutions.}. 

In standard (complex) quantum mechanics, multiplying by complex numbers the 
following sixteen real matrices
\[ \bamq{1}{\cdot}{\cdot}{\cdot} \cdot & \cdot & \cdot & \cdot \\
\cdot & \cdot & \cdot & \cdot \\ \cdot & \cdot & \cdot & \cdot \eam \; , \;
\bamq{\cdot}{1}{\cdot}{\cdot} \cdot & \cdot & \cdot & \cdot \\
\cdot & \cdot & \cdot & \cdot \\ \cdot & \cdot & \cdot & \cdot \eam \; , \;
... \; , \;
\bamq{\cdot}{\cdot}{\cdot}{\cdot} \cdot & \cdot & \cdot & \cdot \\
\cdot & \cdot & \cdot & \cdot \\ \cdot & \cdot & 1 & \cdot \eam \; , \;
\bamq{\cdot}{\cdot}{\cdot}{\cdot} \cdot & \cdot & \cdot & \cdot \\
\cdot & \cdot & \cdot & \cdot \\ \cdot & \cdot & \cdot & 1 \eam \; , \;
\]
(where dots indicate zeros) we obtain the most general 
$4\times 4$ complex matrix, so such a matrix can 
be sufficient to represent the sixteen quantities which characterize the 
Dirac algebra.

At first glance it seems that, within QQM, we must have a Dirac algebra on 
{\em reals} (in we need coefficients $\ag_{A}$ which commute with our 
quaternionic matrices). Utilizing real numbers as multiplicative coefficients, 
we can understand the {\em reduced} dimensions of the $\cg^{\mu}$-matrices, 
because the four {\em real} matrices
\[ \bamd{1}{\cdot}{\cdot}{\cdot} \q , \q 
\bamd{\cdot}{1}{\cdot}{\cdot} \q , \q 
\bamd{\cdot}{\cdot}{1}{\cdot} \q , \q 
\bamd{\cdot}{\cdot}{\cdot}{1} \q 
\]
(if multiplied by real numbers) require the following quaternionic partners
\[ {\bf Q} \bamd{1}{\cdot}{\cdot}{\cdot} \q , \q 
{\bf Q} \bamd{\cdot}{1}{\cdot}{\cdot} \q , \q 
{\bf Q} \bamd{\cdot}{\cdot}{1}{\cdot} \q , \q 
{\bf Q} \bamd{\cdot}{\cdot}{\cdot}{1} \q .
\]
Therefore, working with $2\times 2$ quaternionic matrices and using real 
numbers as multiplicative coefficients we can yet reproduce the {\em magic} 
number 16.

In our precedent papers~\cite{del1,del2,del3,del4,del5,del6} 
we have emphasized the possibility 
to use {\em barred} quaternions within quaternionic matrices. In this case we 
could multiply our matrices by {\em complex} numbers like
\[ a+b\mid i \qq (a, \; b \in {\cal R}) \q , \]
which obviously commute with any quaternionic quantities. If we allow of 
using such barred-complex numbers, we can generalize the 
standard quaternionic trace definition
\[ \mbox{tr} \bamd{q_{1}}{q_{2}}{q_{3}}{q_{4}} = \mbox{re}(q_{1}+q_{4}) \q ,\]
by
\be
\mbox{Tr} \bamd{q_{1}+p_{1}\mid i}{q_{2}+p_{2}\mid i}{q_{3}+p_{3}\mid i}
{q_{4}+p_{4}\mid i} = \mbox{re}(q_{1}+q_{4}) + \mbox{re}(p_{1}+p_{4}) \mid i
\q .
\ee
It is straightforward to prove that the new trace definition guarantees the 
standard property
\[ \mbox{Tr}(M_{1}M_{2})=\mbox{Tr}(M_{2}M_{1}) \q . \]

Choosing {\em barred} complex coefficients $\ag_{A}$ and generalizing the 
trace definition, we can easily demonstrate the theorems 2, 3, 4, given in 
the previous section.

In order to complete our discussion concerning the quaternionic Dirac 
algebra we must modify the Frobenius and Shur theorem as follows
\be n=4n_{1}^{2}+4n_{2}^{2}+ ... +4n_{p}^{2} \q , \ee
in fact we must remember that for any real matrix
\[ \left( \ba{ccc} 1 & 0 & \cdots\\
0     &    0   & \cdots\\
\vdots & \vdots  & \ddots \eam \q ,\]
we must add three quaternionic partners
\[ {\bf Q} \left( \ba{ccc} 1 & 0 & \cdots\\
0     &    0   & \cdots\\
\vdots & \vdots  & \ddots \eam \q .\]

{\bf Modified Frobenius-Schur Theorem}: Let $\cal A$ be an algebra of order $n$ 
possessing a unit element. Let $p$ be the number of (non-equivalent) 
irreducible representations of the algebra, and denote the dimensionality 
of these representations by $n_{1}, \; n_{2}, \; ... \; ,n_{p}$ in turn. 
Then
\bml
\be 
n=4(n_{1}^{ \; 2}+n_{2}^{ \; 2}+ \; ... \; +n_{p}^{ \; 2})  \q , 
\ee
with
\be 
n_{1, \; ..., \; p} ~=~ \frac{1}{2}, \; 1, \; \frac{3}{2}, \; 2, 
\; \frac{5}{2}, \; ... \q . 
\ee
\eml

\section{Quaternionic DKP Algebra}

Applying the {\em modified} Frobenius-Schur theorem to the Dirac algebra we 
find\f{Note that the maximum number of Dirac algebra base elements which 
commute with each other is one, so $n=4n_{1}^{2}$}
\be 16=4n_{1}^{2} \q , \ee
thus we have $2\times 2$ quaternionic matrix representations for the Dirac 
algebra [see eq.~(\ref{rm})].

In this section we briefly show an example of {\em half-whole dimensions} 
by analyzing the DKP algebra
\be \bg^{\mu}\bg^{\nu}\bg^{\lgg}+\bg^{\lgg}\bg^{\nu}\bg^{\mu}=
-g^{\mu \nu}\bg^{\lgg} - g^{\lgg \nu} \bg^{\mu} \q . \ee

By a similar procedure as was used in the case of the Dirac algebra, but of 
course with considerably more effort, one can trace the properties of the 
DKP algebra. We find that there are now 126 linearly independent quantities. 
Moreover, one finds that there are three elements which commute with all 
base elements (p=3). We now may use our {\em modified} theorem as given in 
the previous section. We have to decompose 126 into the sum of three square 
numbers. This is accomplished by
\be 126=4 \left[ \left( \frac{1}{2} \right)^{2} + 
\left( \frac{5}{2} \right)^{2} +  5^{2} \right] \q . \ee
In summary, the DKP algebra has three quaternionic representations and these 
are one-half (trivial), five-half (spin 0) and five (spin 1) 
dimensional representations. 

We explicitly give the quaternionic representations of dimension 
$\frac{1}{2}$ and $\frac{5}{2}$ by the following $3\times 3$ quaternionic 
matrices:
\beal{28}
\bg^{0} = \bamt{\cdot}{\cdot}{a}{\cdot}{\cdot}{\cdot}{-a}{\cdot}{\cdot}
 & \q , \q  & \bg^{1} = j \; 
\bamt{\cdot}{\cdot}{a}{\cdot}{\cdot}{\cdot}{-d}{\cdot}{\cdot} \q , \nonumber \\
 & & \\
\bg^{2} = \bamt{\cdot}{\cdot}{\cdot}{\cdot}{\cdot}{a}{\cdot}{a}{\cdot}
 & \q , \q & \bg^{3} = j \; 
\bamt{\cdot}{\cdot}{\cdot}{\cdot}{\cdot}{a}{\cdot}{-d}{\cdot} \q , \nonumber 
\eea
where
\[ 2a=\frac{1-i\mid i}{2} \q , \q 2d=\frac{1+i\mid i}{2} \q . \]
We can immediatly observe that the $\beta^{\mu}$-matrices of 
eq.~(\ref{28}) act 
trivially on the state
\[ \bat{0}{0}{j} \qq [~\mbox{trivial case}~] , \]
and non-trivially on the states
\[ \bat{1}{0}{0} \q , \q \bat{j}{0}{0} \q , \q \bat{0}{1}{0} \q , \q 
\bat{0}{j}{0} \q , \q \bat{0}{0}{1} \qq [~\mbox{spin 0}~] \q . \]
Such matrices represent the quaternionic counterpart of the complex matrices 
(spin 0 $+$ trivial case) which appear in the standard  DKP equation 
(a complete discussion of the quaternionic DKP equation is recently appeared 
in literature~\cite{del3}).

\section{Conclusions}
The renewed interest in $QQM$ (book~\cite{adl}, and ref.~\cite{adl1}), 
suggests us to look at the 
quaternionic world with trust. The introduction of barred quaternions
\[ q+ p \mid i \q , \]
(natural objects when one works within $QQM_{cg}$) allow us to formulate in 
a consistent way the standard physical theories (like special 
relativity~\cite{del4}, electroweak model~\cite{del5}, GUT~\cite{del6}). 
From the viewpoint of group structure, these barred quantities are very 
similar to complexified quaternions~\cite{mor}
\[ q + {\cal I}p \q  \]
(the imaginary unit $\cal I$ commutes with the quaternionic imaginary units 
$i, \; j, \; k$), but in physical problems, like eigenvalue calculations, 
tensor products, relativistic equations solutions, they give different 
results.

Barred quaternions are very useful in writing a quaternionic version of 
the Dirac~\cite{rot} and DKP~\cite{del3} equations. Nevertheless, if we 
wish to use quaternions as underlying numerical field we must revise the   
standard assumptions. For example, due to the doubling of solutions given by 
the complex geometry, we have to introduce {\em half-whole} 
dimensions for quaternionic matrices and modify the Frobenius-Schur 
theorem. Obviously, this represents only a first step towards a 
quaternionic world. An interesting research topic could be to generalize 
the group theoretical structure by our barred quaternionic operators.

\bref
\bi{adl}
S.~L.~Adler, {\it Quaternionic Quantum Mechanics and Quantum Fields} 
(Oxford, New York, 1995).
\bi{hor}
L.~P.~Horwitz and L.~C.~Biedenharn, \axp{157}{432}{84}. 
\bi{rem}
J.~Rembieli\'nski, \jxg{11}{2323}{78}.
\bi{rot}
P.~Rotelli, \mxb{4}{933}{89}.
\bi{del1}
S.~De Leo and P.~Rotelli, \pxxa{92}{917}{94}.
\bi{del2}
S.~De Leo and P.~Rotelli, {\it Odd Dimensional Translation between Complex 
and Quaternionic Quantum Mechanics}, 
Prog.~Theor.~Phys. (submitted).
\bi{del3}
S.~De Leo, \pxxa{94}{11}{95}.
\bi{del4}
S.~De Leo, {\it Quaternions and Special Relativity}, J.~Math.~Phys. 
(to be published, June 1996).
\bi{del5} 
S.~De Leo and P.~Rotelli, \ixa{10}{4359}{95}.\\
S.~De Leo and P.~Rotelli, {\it Quaternionic Dirac Lagrangian}, 
Mod.~Phys.~Lett.~A (to be published).\\
S.~De Leo and P.~Rotelli, {\it Quaternionic Electroweak Theory}, 
J.~Phys.~G (submitted).
\bi{del6}
S.~De Leo, {\it Quaternions for GUTs}, Int.~J.~Theor.~Phys. (submitted).
\bi{adl1}
S.~L.~Adler,  \nxb{B415}{195}{94}; \pxi{332B}{358}{94}.
\bi{mor}
K.~Morita, \pxxa{67}{1860}{81}; \xxx{68}{2159}{82}; \xxx{70}{1648}{83}; 
\xxx{72}{1056}{84}; \xxx{73}{999}{84}; \xxx{75}{220}{85}; \xxx{90}{219}{93}. 

\eref

%==============================================================================
\end{document}